\documentclass[conference, letterpaper]{IEEEtran} 
\IEEEoverridecommandlockouts

\usepackage{amsmath, amssymb, cite}
\usepackage{amsthm,bm,bbm}
\usepackage{mathtools}
\usepackage[size=small]{caption}
\usepackage{amsthm,multirow,color,amsfonts}
\usepackage{tabulary}
\usepackage{subfigure}
\usepackage{graphicx}
\usepackage{setspace}
\usepackage{enumerate}
\usepackage{algorithm}
\usepackage{algpseudocode}
\usepackage{comment}
\usepackage{pdfpages}
\usepackage{hyperref}
\usepackage{stfloats}

\usepackage{etoolbox}
\makeatletter
\patchcmd{\@makecaption}
  {\scshape}
  {}
  {}
  {}
\makeatother

\makeatletter
\def\BState{\State\hskip-\ALG@thistlm}
\makeatother

\usepackage[utf8]{inputenc} 
\usepackage[T1]{fontenc}
\usepackage{url}
\usepackage{ifthen}

\usepackage[dvipsnames]{xcolor}

\newtheorem{theo}{Theorem}
\newtheorem{defi}{Definition}

\newtheorem{prop}{Proposition}

\newtheorem{ex}{Example}

\newcommand{\err}{\mathcal{E}r}
\newcommand{\Rcomp}{R'}

\newcommand{\GU}{G_{\cup_{i=1}^{K}f_i}}

\def\compactify{\itemsep=0pt \topsep=0pt \partopsep=0pt \parsep=0pt}
\let\latexusecounter=\usecounter

\ifx\du\undefined
  \newlength{\du}
\fi
\usepackage{siunitx}
\usepackage{tikz}
\usetikzlibrary{shapes.geometric}
\usepackage{verbatim}
\usepackage{color,soul}
\usetikzlibrary{spy,calc}
\hypersetup{breaklinks=true}
\usepackage{diagbox}

\tikzstyle{vertex}=[circle,minimum size=1pt,inner sep=0pt]
\tikzstyle{selected vertex} = [vertex, fill=red!24]
\tikzstyle{edge} = [draw,thick,-]
\tikzstyle{weight} = [font=\small]
\tikzstyle{selected edge} = [draw,line width=5pt,-,red!50]
\tikzstyle{ignored edge} = [draw,line width=5pt,-,black!20]

\begin{document}
\title{Non-Linear Function Computation Broadcast

\thanks{Co-funded by the European Union (ERC, SENSIBILITÉ, 101077361). Views and opinions expressed are however those of the authors only and do not necessarily reflect those of the European Union or the European Research Council. Neither the European Union nor the granting authority can be held responsible for them. 

This research was partially supported by a Huawei France-funded Chair towards Future Wireless Networks, and supported by the program "PEPR Networks of the Future" of France 2030.}
\thanks{
The first two authors have equally contributed to this work.}
} 
\author{
   \IEEEauthorblockN{Mohammad Reza Deylam Salehi, Vijith Kumar Kizhakke Purakkal, Derya Malak}
   \IEEEauthorblockA{Communication Systems Department, EURECOM, Biot Sophia Antipolis, France\\
   \{deylam, kizhakke, malak\}@eurecom.fr}}

\maketitle

\begin{abstract}
This work addresses the $K$-user computation broadcast problem consisting of a master node, that holds all datasets and users for a general class of function demands, including linear and non-linear functions, over finite fields. 
The master node sends a broadcast message to enable each of $K$ distributed users to compute its demanded function in an asymptotically lossless manner with user's side information. 
We derive bounds on the optimal $K$-user computation broadcast rate that allows the users to compute their demanded functions by capturing the structures of the computations and available side information.
Our achievability scheme involves the design of a novel graph-based coding model to build a broadcast message to meet each user's demand, by leveraging the structural dependencies among the datasets, the user demands, and the side information of each user, drawing on K{\"o}rner's characteristic graph framework. 
The converse uses the structures of the demands and the side information available at $K$ users to yield a tight lower bound on the broadcast rate. 
With the help of examples, we demonstrate our scheme achieves a better communication rate than the existing state of the art. 
\end{abstract}



\section{Introduction}
\label{sec:intro}
In the presence of large-scale data, communication networks are being increasingly deployed for various distributed computation scenarios, including distributed learning\cite{predd2006distributed, 10129894}, over-the-air computing~\cite{csahin2023survey, montalban2021broadcast}, private-key encryption~\cite{kate2010distributed, liu2020privacy}, and coded distributed computing\cite{li2016coded, dutta2019optimal}, to name a few. 
In such scenarios, meeting the growing and diverse demands of distributed users is exacerbated by the limited communication resources, privacy constraints, and bounded memory and computational capabilities of users. 
This motivates us to study the problem of computation broadcast, in which source data is communicated to recover functions of data at distributed users, which finds applications in radio broadcasting, digital video broadcasting, and cryptography~\cite{chlebus2000deterministic, reimers1998digital, cryptoeprint:2024/154}.

\subsection{Motivation and Contributions}
\label{sec:motivation}
 
Prior works in computation broadcast have focused on linear function demands~\cite{ wan2021optimal,sun2019capacity, yao2024capacity, yao2024generic}, which allow for exploiting algebraic techniques to encode the source data, or complementary function demands~\cite{ravi2016broadcast}. However, these approaches often depend on the specific demand structures or the number of users, limiting their general applicability. Furthermore, to the best of our knowledge, no general bound on the broadcast rate has been established for non-linear demands.

This paper tackles the problem of function computation broadcast, accommodating general function demands (linear, non-separable, and non-linear) over finite fields. We develop a master-users framework using {\emph{characteristic graphs}} to capture demand structures. The master node, has access to all datasets and computation capabilities, transmits information to multiple users. Each user, having a fraction of the master's datasets as side information, computes a general function of the datasets.

Our main contributions are summarized as follows.
\begin{itemize}
    \item We devise a coding scheme for the $K$-user computation broadcast that tackles general data distributions, and general non-linear function demands over $\mathbb{F}_q$, denoting a finite field of order $q\geq 2$. 
    This framework, as detailed in  Section~\ref{sec:sys-mod}, emerges from using characteristic graphs devised by K\"orner~\cite{korner1973coding} to identify a collection of equivalent source classes. The master builds a broadcast message to enable all users to decode their demanded functions simultaneously in an asymptotically lossless manner.

    \item This framework reduces communication costs by performing part of the computation at the master node and transmitting functions of datasets as the broadcast message, incorporating the structure of datasets and $K$ user demands (Sections~\ref{sec:sys-mod} and~\ref{sec:3-user-comp-broad}). For general non-linear demands, our approach outperforms~\cite{SlepWolf1973}. In a $3$-user Boolean computation broadcast model, the achievable rate is $1.5$ bits, improving on the $2$ bits of~\cite{SlepWolf1973} (Example~\ref{ex-nonlinear-comp}, Section~\ref{sec:3-user-comp-broad}). Similarly, for a $3$-user linear computation broadcast model, we achieve $1.42$ bits, outperforming~\cite{yao2024capacity} and~\cite{SlepWolf1973} with rates of $1.5$ and $2$ bits, respectively (Example~\ref{ex-linear-comp-jafar}, Section~\ref{sec:3-user-comp-broad}).

     \item  We derive an achievability scheme and a converse on the communication rate for $K\geq 2$-user computation broadcast with linear and non-linear demands in Section~\ref{sec:main-resu}. 
\end{itemize}

\subsection{State of the Art}
\label{sec:SoA}
\textbf{Communication complexity:} 
Communication complexity, which quantifies the minimum bits required for transmission among multiple parties, was introduced for distributed computation of Boolean functions~\cite{yao1979some} and later extended to approximating continuous functions~\cite{tsitsiklis1987communication}, computing rational functions~\cite{luo1989communication}, and estimating correlations~\cite{hadar2019communication}. Interactive communication models have demonstrated significant communication savings compared to one-way models~\cite{ma2012interactive, ma2011some}. While previous studies have explored communication complexity across various multi-party communication models, characterizing this complexity for computation broadcast problems with general function demands remains an open problem.

\textbf{Compression for computing:}
Structured codes reduce rates in distributed computation for tasks like modulo-$q$ sum~\cite{korner1979encode,lalitha2013linear}, linear functions~\cite{nazer2007computation}, and matrix multiplication~\cite{malak2024structured}. Unstructured codes address point-to-point lossy recovery~\cite{WynZiv1976}, function computation with side information~\cite{Yamamoto1982}, and multi-terminal setups~\cite{SlepWolf1973,tuncel2006slepian,gray1974source,li2017extended}. K{\"o}rner's characteristic graph framework~\cite{korner1973coding} is used to compress sources for distributed computing~\cite{han1987dichotomy, AlonOrlit1996, OrlRoc2001, feizi2014network, malak2024multis, salehi2023achievable, malak2023distributed,malak2024multi, ozyilkan2024distributed}. Extending this to broadcast models is difficult due to the need to handle function and side information dependencies in message design.

\textbf{Coded distributed computing}:
To meet growing computational demands and speed up computing via distributing tasks among multiple servers, techniques like distributed machine learning~\cite{lee2017speeding}, federated learning~\cite{cui2025toward}, and massive parallelization algorithms, e.g., MapReduce~\cite{dean2008mapreduce} and Spark~\cite{zaharia2010spark}, have been proposed. 
Most strategies exploit the separable nature of functions, e.g., matrix multiplication~\cite{jia2021capacity} and gradient descent~\cite{tandon2017gradient}, and design codes to distribute tasks, focusing on straggler mitigation~\cite{yu2020straggler}, communication complexity~\cite{tandon2017gradient,dutta2019optimal}, and communication-computation tradeoffs~\cite{khalesi2022multi, jia2021capacity}. However, they fall short for general non-linear demands.

\textbf{Computation broadcast:} 
The computation broadcast framework generalizes the index coding problem~\cite{arbabjolfaei2018fundamentals}, where users demand source variables instead of their functions, and multiple description coding~\cite{gamal1982achievable}, a special case where all users demand the same distorted source function over $\mathbb{F}_q$ with varying distortion criteria. Linear computation broadcast has been studied for two and more users settings\cite{yao2024capacity,yao2024generic}. However, for non-linear computation broadcast, only the case of $K=2$ users with complementary side information has been explored~\cite{ravi2016broadcast}. Here, we extend \cite{ravi2016broadcast} to $K\geq 2$ users without structural assumptions on side information.

\section{System Model}
\label{sec:sys-mod}
\begin{figure}[t!]
\centering
\input{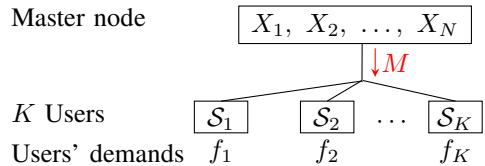}
\caption{Computation broadcast model for $K$ users, each with side information $\mathcal{S}_i$ and demand $f_i$, for $i\in [K]$ \cite{sun2019capacity}.}
\label{fig:CBNs}
\end{figure}

This section outlines our $K$-user computation broadcast model, depicted in Figure \ref{fig:CBNs}. The model consists of a master node, that stores $N\in\mathbb{Z}^+$ datasets, denoted by $X_{[N]}\triangleq~\{X_j\}_{j\in[N]}$, where $X_j \in \mathbb{F}_q$ for $j\in[N]\triangleq\{1,\cdots,N\}$ and $q \geq 2$, and $K$ users.
Each of $[K]$ distributed users is equipped with side information, $\mathcal{S}_i\subseteq X_{[N]}$. Users are interested in computing a set of (possibly distinct) functions $\{f_i\}_{i\in [K]}$ using the shorthand notation $f_i=f_{i}(X_1, \dots, X_N)$. 

The master node, using $n$ i.i.d. realizations of 
${X}^n_j=\{X_{j,t}\}_{t\in [n]}\in \mathbb{F}_{q}^{n}$, broadcasts a message $M=\psi(X^n_1,\dots, X^n_N)$, designed as a function of the master node's data and the set of demanded functions. The objective of the computation broadcast problem is to minimize the communication rate from the master node, characterized by the entropy of $M$, denoted by $H_q(M)=\mathbb{E}[- \log_{q}P_{M}]$, that ensures each user $i\in[K]$ to asymptotically losslessly reconstruct the length $n$ sequence of functions $f_i^n\triangleq\{f_{i}({X}_{1,t}, \cdots, X_{N,t})\}_{t\in[n]}$, using the received message $M$ and  the length $n$ realization of $\mathcal{S}_i$, i.e.,  $\mathcal{S}_{i}^{n}=\{\mathcal{S}_{i,t}\}_{t\in[n]}$, i.e., the following condition is met:
\begin{equation}
\label{Recovery_condition}
    H_q(f_{i}^n\mid M, \mathcal{S}^n_{i})=0\ , \quad \forall i\in [K] \ .
\end{equation}

To capture general function demands in the proposed $K$-user computation broadcast model, we exploit the characteristic graph framework in~\cite{korner1973coding}, which has been used to characterize the communication rate for a point-to-point model, where a source holds a random variable $X$ and a receiver has side information $Y$, and the receiver's goal is to reconstruct a function $f(X, Y)$ in an asymptotically lossless manner~\cite{OrlRoc2001}. To motivate our approach, we next define characteristic graphs.

\begin{defi}[{\bf Source characteristic graphs} \cite{OrlRoc2001}]
\label{def-char-graph}
 Consider a point-to-point 
 model with a master node having access to a 
 source random variable $X\in\mathcal{X}$ and a user with side information 
 $Y\in\mathcal{Y}$, aiming to compute $f(X, Y)$. The characteristic graph that the master node builds using $X$ for $f(X, Y)$, given 
 $Y$, is defined as $G_f=G(\mathcal{V}, \mathcal{E})$, where  $\mathcal{V}=\mathcal{X}$ denotes the vertex set
 , and $\mathcal{E}$ the set of edges such that given two distinct vertices $x, x' \in \mathcal{V}$, an edge $(x, x') \in \mathcal{E}$ exists if and only if the following conditions hold\footnote{Given random variables 
 $X$ and $Y$, the notation $P_{X}$ denotes the probability mass function (PMF) of $X$, and $P_{X, Y}$ denotes the joint PMF of $X$ and $Y$.}: i) $f(x,y) \neq f(x',y)$ and ii) $P_{X, Y}(x,y)\cdot P_{X, Y}(x',y)>0$ for some $y\in \mathcal{Y}$. 
\end{defi}

Characteristic graphs identify which source values can share the same codeword without introducing ambiguity to the user, enabling distributed computation in various topologies, as explored by prior literature, see e.g.,~\cite{AlonOrlit1996, OrlRoc2001,feizi2014network, salehi2023achievable, malak2024multi, malak2024multis,korner1973coding,malak2023distributed}. For the point-to-point scenario detailed above, the minimum rate required for a user to compute $f(X, Y)$ given side information $Y$ with vanishing error probability is characterized as~\cite{OrlRoc2001}
\begin{equation}
\label{eq:Rate_ach}
    \min_{\substack{W-X-Y\\X\in W\in \Gamma(G_f)}} I(X;W\mid Y) \ ,
\end{equation}
where $W-X-Y$ indicates a Markov chain,  $I(X; W \mid Y)=H_{q}(W\mid Y)-H_{q}(W\mid X, Y)$, and  $X\in W\in \Gamma(G_f)$ means that the minimization is over all joint PMFs $P_{W, X}(w,x)>0$ such that $W$ is an independent set\footnote{An independent set (IS) is a subset of vertices in $G_f$ where no two vertices are connected by an edge, ensuring that they can be encoded without ambiguity. A maximal independent set (MIS), denoted by $W\in \Gamma(G_f)$, is an independent set that cannot be extended by including additional vertices~\cite{beigel1999finding}.} of $G_f$, and $\Gamma(G_f)$ is the collection of all MISs of $G_f$.  

The paper extends the characteristic graph framework to address the computation broadcast problem by designing a broadcast message, $M$, that captures dependencies among the master's datasets $X_{[N]}$, the structure of $f_i$, and each user's side information $\mathcal{S}_i$ for $i \in [K]$. This approach serves all users' demands simultaneously without imposing structural constraints (e.g., linearity, separability), unlike prior works~\cite{sun2019capacity,ravi2016broadcast,yao2024capacity,yao2024generic}. To jointly capture the distinguishability requirements of $K$ users, the framework leverages union graphs, as defined next.

\begin{figure*}[t!]
    \centering
    \begin{minipage}{0.24\textwidth}
        {\begin{tikzpicture}[scale=0.8, auto,swap]
            \foreach \pos/\name/\value/\colo/\labelpos in {{(2,3)/a/000/black!30!white/above}, {(3,2)/b/001/black!30!white/above}, {(3,1)/c/010/black!30!white/below}, {(2,0)/d/011/black!30!white/below}, {(1,0)/e/100/black!30!white/below}, {(0,1)/f/101/black!30!white/below},{(0,2)/g/110/black!30!white/above},{(1,3)/h/111/black!30!white/above}}
            \node[draw, circle, inner sep=2pt, fill=black, label={[label distance=1mm]\labelpos:$\value$}](\name) at \pos {};
            \foreach \source/ \dest /\weight in {h/e/,h/f/, h/g/ }
                \path[edge, color=black] (\source) [dash pattern=on \pgflinewidth off 2pt]--node[weight] {$\weight$} (\dest);
        \end{tikzpicture}}
    \end{minipage}%
    \begin{minipage}{0.24\textwidth}
        {\begin{tikzpicture}[scale=0.8, auto,swap]
            \foreach \pos/\name/\value/\colo/\labelpos in {{(2,3)/a/000/black!30!white/above}, {(3,2)/b/001/black!30!white/above}, {(3,1)/c/010/black!30!white/below}, {(2,0)/d/011/black!30!white/below}, {(1,0)/e/100/black!30!white/below}, {(0,1)/f/101/black!30!white/below},{(0,2)/g/110/black!30!white/above},{(1,3)/h/111/black!30!white/above}}
            \node[draw, circle, inner sep=2pt, fill=black, label={[label distance=1mm]\labelpos:$\value$}](\name) at \pos {};
            
            \foreach \source/ \dest /\weight in {b/a/, a/f/,a/e/ }
                \path[edge, color=black] [dash pattern=on 4pt off 2pt](\source) -- node[weight, anchor=west] {$\weight$} (\dest);
        \end{tikzpicture}}
    \end{minipage}%
    \begin{minipage}{0.24\textwidth}
        {\begin{tikzpicture}[scale=0.8, auto,swap]
            \foreach \pos/\name/\value/\colo/\labelpos in {{(2,3)/a/000/black!30!white/above}, {(3,2)/b/001/black!30!white/above}, {(3,1)/c/010/black!30!white/below}, {(2,0)/d/011/black!30!white/below}, {(1,0)/e/100/black!30!white/below}, {(0,1)/f/101/black!30!white/below},{(0,2)/g/110/black!30!white/above},{(1,3)/h/111/black!30!white/above}}
            \node[draw, circle, inner sep=2pt, fill=black, label={[label distance=1mm]\labelpos:$\value$}](\name) at \pos {};
            
            \foreach \source/ \dest /\weight in {h/b/, h/d/, f/b/, f/d/ }
                \path[edge] (\source) -- node[weight, anchor=east] {$\weight$} (\dest);
            \foreach \source/ \dest /\weight/\coloedge in {a/g/, g/e/,g/c/ }
                \path[edge, color=black] [dash pattern=on 4pt off 4pt](\source) -- node[weight] {$\weight$} (\dest);
        \end{tikzpicture}}
    \end{minipage}
    \begin{minipage}{0.24\textwidth}
        \begin{tikzpicture}[scale=0.8, auto,swap]
            \foreach \pos/\name/\value/\colo/\labelpos in {{(2,3)/a/000/black!30!white/above}, {(3,2)/b/001/black!30!white/above}, {(3,1)/c/010/black!30!white/below}, {(2,0)/d/011/black!30!white/below}, {(1,0)/e/100/black!30!white/below}, {(0,1)/f/101/black!30!white/below},{(0,2)/g/110/black!30!white/above},{(1,3)/h/111/black!30!white/above}}
            \node[draw, circle, inner sep=2pt, fill=black, label={[label distance=1mm]\labelpos:$\value$}](\name) at \pos {};
            \foreach \source/ \dest /\weight in {h/e/,h/f/, h/g/ }
                \path[edge, color=black] (\source) [dash pattern=on \pgflinewidth off 2pt]--node[weight] {$\weight$} (\dest);
            
            \foreach \source/ \dest /\weight in {g/a/, g/c/,g/e/ }
                \path[edge, color=black] [dash pattern=on 4pt off 4pt](\source) -- node[weight, anchor=west] {$\weight$} (\dest);

            \foreach \source/ \dest /\weight in {h/b/, h/d/, f/b/, f/d/ }
                \path[edge] (\source) -- node[weight, anchor=east] {$\weight$} (\dest);
            \foreach \source/ \dest /\weight/\coloedge in {a/b/, a/e/,a/f/ }
                \path[edge, color=black] [dash pattern=on 4pt off 2pt](\source) -- node[weight] {$\weight$} (\dest);
        \end{tikzpicture}
    \end{minipage}
    \caption{Individual characteristic graphs and broadcast graph for Example~\ref{ex-nonlinear-comp}: (Left)  $G_{f_1}$, (Middle left) $G_{f_2}$, and (Middle right) $G_{f_3}$, based on users' demands $f_1$, $f_2$, and $f_3$, and the side information $\mathcal{S}_i$ for each $i\in[3]$, respectively, and (Right) broadcast graph $G_{\cup_{i=1}^{3}f_{i}}$.}
    \label{fig:Graphs}
\end{figure*}
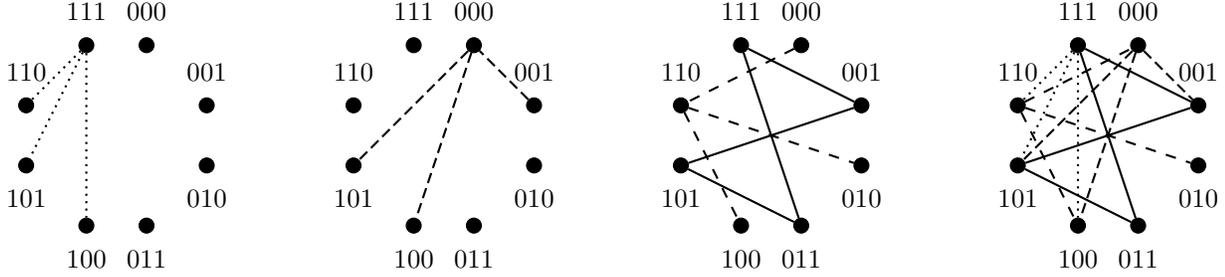

\begin{defi}[{\bf Union graph}\cite{korner1986fredman}]
\label{def:union_graph}
Given two characteristic graphs  $G_{f_1} =G(\mathcal{V}, \mathcal{E}_1)$ and $G_{f_{2}}=G(\mathcal{V}, \mathcal{E}_2)$ with the same vertex set $\mathcal{V}$, their union graph,  $G_{f_{1}\cup f_2}$, is defined as:
\begin{equation}
\label{Union_graph_def_eq}
    G_{f_{1}\cup f_2} \triangleq G(\mathcal{V}, \mathcal{E}_{1}\cup \mathcal{E}_{2})\ ,
\end{equation}
where the set of edges is given by the union $\mathcal{E}_{1}\cup \mathcal{E}_{2}$, which captures if a pair of source values need to be distinguished for at least one of the two functions $f_1$ and $f_2$.
\end{defi}

Next, using the union graph approach, we will devise an achievability scheme for the computation broadcast problem, which involves the design of the broadcast message $M$. Generalizing Definition~\ref{def:union_graph}, we denote by $\GU$ the union graph --- referred to as broadcast graph --- built by the master node to simultaneously meet the demands of all $K$ users where $\Gamma(\GU)$ is the set of all MISs of this broadcast graph. From Definition~\ref{def:union_graph}, an MIS $W$ of $\GU$ satisfies the constraint (\ref{Recovery_condition}). 

Assigning a unique codeword to the MISs in $\Gamma(\GU)$ ensures that no confusion arises by the assigned codeword in any user $i\in[K]$ in reconstructing $f_{i}$. 
For asymptotically lossless recovery of all demands, the master node builds the $n$-th OR power\footnote{The $n$-fold OR products of characteristic graphs are detailed in \cite{AlonOrlit1996}.} $G^n_{\cup_{i=1}^{K} f_i}$ corresponding to the length $n$ realization of $X_{[N]}$. 
To that end, the master node uses a binning strategy\footnote{Theorem~\ref{theo-inner-and-outer} in Section~\ref{sec:main-resu} details the binning strategy that achieves $R_{ach}$.} and builds a mapping $en:\Gamma(G^n_{\cup_{i=1}^{K} f_i})\to [0, q^{nR_{ach}}-1]$,  
to construct the broadcast message $M=en(W^n)$ using the MISs $W^n\in\Gamma(G^n_{\cup_{i=1}^{K} f_i})$, such that user $i\in [K]$ computes the non-componentwise function $\hat{f}_{i}^{n}\triangleq F_i(W^n, \mathcal{S}_{i}^{n})$, and the condition $P(\hat{f}_{i}^{n}\neq f_{i}^{n})<\epsilon$ is ensured asymptotically for some $\epsilon>0$, rendering the condition in (\ref{Recovery_condition}) true for $M$.

Hence, generalizing (\ref{eq:Rate_ach}) to the broadcast setting, and using Slepian-Wolf coding~\cite{SlepWolf1973}, the broadcast rate 
satisfies
\begin{equation}
\label{eq-com-rate-main-mutual-info}
 R_M\leq\min_{X_{[N]}\in W\in \Gamma(G_{\cup_{i'=1}^{K}f_{i'}})}\max_{i\in [K]} I(X_{[N]} ;W\mid \mathcal{S}_{i}) \ ,  
\end{equation}
where the minimization is over all $P_{W, X_{[N]}}(w,x_{[N]})>0$ such that $W\in\Gamma(\GU)$ and (\ref{Recovery_condition}) is ensured\footnote{The right hand side in (\ref{eq-com-rate-main-mutual-info}) is achievable by assigning colors to 
$G^n_{\cup_{i=1}^{K} f_i}$ which achieves the entropy bound of the coloring random variable~\cite{korner1973coding,feizi2014network}.}. Since $W$ is a function of $X_{[N]}$, the following substitution holds in (\ref{eq-com-rate-main-mutual-info}):
\begin{equation}
\label{eq-rate-mutual-convert-entropy}
    I(X_{[N]} ;W\mid \mathcal{S}_{i})=H_q(W\mid \mathcal{S}_{i}) \ .  
\end{equation}


We next demonstrate our coding scheme through examples.

\section{\texorpdfstring{$3$}{}-user Computation Broadcast: Examples}
\label{sec:3-user-comp-broad}
Next, we focus on a $3$-user computation broadcast model, with $N=3$ datasets $X_{[3]}$, where each $X_{j}$ for $j\in[3]$ is i.i.d. and uniformly distributed over $\mathbb{F}_q$. Each user $i\in[3]$ has 
$\mathcal{S}_i=X_i$ and requests a distinct function $f_i=f_i(X_{1},X_{2},X_{3})$. The former example examines Boolean (non-linear) function demands, while the latter focuses on linear demands.

\begin{ex}[\textbf{\texorpdfstring{$3$}{}-user Boolean computation broadcast}]
\label{ex-nonlinear-comp}
We denote user requests by $f_1=X_1\wedge X_2\wedge X_3$, $f_2=X_1\vee X_2\vee X_3$, and $f_3=X_1\wedge(X_2\vee X_3)$, where $X_{j}\in\mathbb{F}_2$ for $j\in[3]$, $\mathcal{S}_{i}=X_{i}$ for $i\in [3]$, and $\wedge$ and $\vee$ denote binary AND and OR operations, respectively. 

The characteristic graph $G_{f_{1}}=(\mathcal{V},\mathcal{E}_1)$ 
given $\mathcal{S}_1=X_1$ is shown in Figure~\ref{fig:Graphs} (Left), where $\mathcal{V}=\mathbb{F}_2\times\mathbb{F}_2\times\mathbb{F}_2$. For $X_1=0$, the outcome of $f_1 =0$ for any $(X_2,X_3)$ pair. Thus, in $G_{f_{1}}$, we don't need to distinguish the vertices with $X_1=0$. When $X_1=1$, $f_1 =1$ if and only if $X_2=X_3=1$, and otherwise $f_1=0$. Thus, $\mathcal{E}_1=\{(111,100), (111,101),(111,110)\}$. 
Similarly, we construct $G_{f_{2}}$ and $G_{f_{3}}$, as shown in Figure~\ref{fig:Graphs} (Middle left) and (Middle right), respectively. Using Definition~\ref{def:union_graph},  $G_{\cup_{i=1}^{3}f_i}$ is shown in Figure~\ref{fig:Graphs}~(Right).

Given $G_{\cup_{i=1}^{3}f_i}$ that captures all user demands and side information $\{\mathcal{S}_i\}_{i\in[3]}$, the rate of the broadcast message, employing (\ref{eq-com-rate-main-mutual-info}) and (\ref{eq-rate-mutual-convert-entropy}), is given by 
\begin{equation}
\label{R-3-user}
    R_M\leq\min_{X_{[3]}\in W\in \Gamma(G_{\cup_{i=1}^{3}f_i})} \max_{{\begin{array}{c}
         i\in[3]
    \end{array}}}H_2(W\mid X_{i}) \ ,
\end{equation}
where $G_{\cup_{i=1}^{3}f_i}$ has the following set of MISs:
 \begin{equation}
 \label{Gamma-3-user}
\Gamma(G_{\cup_{i=1}^{3}f_i)}=\left\{\begin{array}{c}
       \{000,010,011\},\{001,110,011\},\\
       \{001,010,011,100\}, \{101,110\},\\ \{ 010,100,101\},
       \{000,010,111\}
  \end{array}\right\}.
  \end{equation}
    
The joint PMF $P_{W, X_{[3]}}$ that minimizes (\ref{R-3-user}) is given in Table~\ref{tab:optimal-joint-distribution}, using which (\ref{R-3-user}) yields $R_{M}\leq1.5$ bits.
\begin{table}[h!]
\setlength{\tabcolsep}{2pt}
    \begin{center}
          \begin{tabular}
          {|c|c|c|c|c|c|c|c|c|}
            \hline&&&&&&&&\\[-1.2em]
            $W\,\backslash\, X_{[3]}$&$000$&$001$&$010$&$011$&$100$&$101$&$110$&$111$\\
             \hline &&&&&&&&\\[-1.em]
             $\left\{
                 000,010,111 \right\}$&$\frac{1}{8}$&$0$&$0$&$0$&$0$&$0$&$0$&$\frac{1}{8}$\\[0.1em]
             \hline&&&&&&&&\\[-1.1em]
             $\left\{
                  001,010,011,100
             \right\}$&$0$&$\frac{1}{8}$&$\frac{1}{8}$&$\frac{1}{8}$&$\frac{1}{8}$&$0$&$0$&$0$  \\[0.2em]
             \hline&&&&&&&&\\[-1.1em]
             $\{101,110\}$&$0$&$0$&$0$&$0$&$0$&$\frac{1}{8}$&$\frac{1}{8}$&$0$\\[0.1em]
             \hline
        \end{tabular}
        \end{center}
\caption{ Joint PMF $P_{W,X_{[3]}}$ that solves (\ref{R-3-user}).}
\label{tab:optimal-joint-distribution}
\end{table}
\end{ex}

Example~\ref{ex-nonlinear-comp} demonstrates that designing a broadcast message leveraging the structure of all demands and users' side information achieves a rate of $R_{ach}=1.5$ bits. In contrast, using $\mathcal{S}_i = X_i$ at $i\in[3]$ and applying the Slepian-Wolf theorem~\cite{SlepWolf1973}, sending $(X_1\oplus_2 X_2, X_2\oplus_2 X_3)$ achieves the minimum rate for asymptotically lossless reconstruction at each user, as $H_2(X_1\oplus_2 X_2, X_2\oplus_2 X_3)=2$ bits~\cite{SlepWolf1973} for all $i\in[3]$, where $\oplus_q$ denotes modulo-$q$ sum. Thus, our graph-based approach achieves a $25\%$ rate reduction versus~\cite{SlepWolf1973}.

Next, we detail a linear computation broadcast example from~\cite{yao2024capacity} and show that our approach outperforms that of~\cite{yao2024capacity}.

\begin{ex}[\textbf{\texorpdfstring{$3$}{}-user linear computation broadcast~\cite[Example~2]{yao2024capacity}}]
\label{ex-linear-comp-jafar}
We denote user requests by $f_{1}=X_{2}\oplus_{3}X_{3}$, $f_2=X_{1}\oplus_{3}X_{3}$, and $f_3=X_{1}\oplus_{3}2X_{2}$, respectively, where $X_{j}\in\mathbb{F}_3$ for $j\in[3]$, and $\mathcal{S}_{i}=X_{i}$ for $i\in [3]$. 

For the proposed setup, the rate required to encode $G_{\cup_{i=1}^{3}f_i}$, i.e., $R_{M}$, is upper bounded by the solution of (\ref{eq-com-rate-main-mutual-info}). However, the broadcast graph $\GU$ has $|\mathcal{V}|=q^N$ vertices, rendering the analytical solution for $\Gamma(\GU)$ and hence for (\ref{eq-com-rate-main-mutual-info}) infeasible for large $q^N$, as compared to Example~\ref{ex-nonlinear-comp} with fewer vertices. 
However, using the sub-additivity properties of graphs, we can divide the solution space and address less complex tasks. 
To that end, we propose an alternative approach for designing $M$ by considering pairs of users and defining compatible functions for each pair, which we detail below.

{\bf Compatibility function-based approach.} To build our achievability scheme, we consider any pair $(i_1,i_2)$ of users $i_1,i_2\in [3]$ with respective demands $f_{i_1}$ and $f_{i_2}$, and with $\mathcal{S}_{i_1}=X_{i_1}$ and $\mathcal{S}_{i_2}=X_{i_2}$, respectively. The rate required to encode $G_{f_{i_{1}}\cup f_{i_{2}}}$, denoted as $R_{i_{1}i_{2}}$, must satisfy 
\begin{align}
    R_{i_{1}i_{2}}\leq \min_{X_{[3]}\in W\in \Gamma(G_{f_{i_{1}}\cup f_{i_{2}}})}\max_{i\in\{i_1,i_2\}}\left\{
    H_q(W\mid X_{i})\right\}\ ,\nonumber
\end{align}
where instead of explicitly determining $\Gamma(G_{f_{i_{1}}\cup f_{i_{2}}})$, we use a compatibility-based approach and demonstrate its optimality for reconstructing $f_{i_1}$ and $f_{i_2}$. 
A given function $Z_{{i_1}{i_2}}$ is called {\emph{a compatible function}} for pair of users $(i_{1},i_{2})$, if it yields $\phi_{i_1}(Z_{{i_1}{i_2}},X_{i_1})=f_{i_1}$ and $\phi_{i_2}(Z_{i_1 i_2},X_{i_2})=f_{i_2}$. For users $(1,2)$, with demands given above, a compatible function is $Z_{12}=X_{1}\oplus_{3} X_{2}\oplus_{3} X_{3}$. Similarly, for user pairs $(1,3)$, and $(2,3)$,  functions $Z_{13}=X_{1}\oplus_{3} 2 X_{2}\oplus_{3} 2X_{3}$ and $Z_{23}=X_{1}\oplus_{3} 2X_{2}\oplus_{3} X_{3}$, respectively, are compatible. For any given $(i_1,i_2)$, the rate achieved by $Z_{{i_1}{i_2}}$ satisfies
\begin{equation}
    R_{i_1i_2}\leq H_3(Z_{{i_1}{i_2}})=H_3(f_{i_1}\mid X_{i_1})=1 \:\text{ bit}, \quad i_1,i_2\in[3] \ .\nonumber
\end{equation}
 
We split a dataset $X_j$ into two disjoint sub-variables $X_j^1$ and $X_j^2$, each with a size of $0.5$ bits, similar to~\cite[Example~2]{yao2024capacity}.   
Let $Z_{{i_1}{i_2}}^l$, for $l\in[2]$, denote the compatible function sub-variables $X_{j}^{l}$, for $l\in[2]$, $j\in[3]$, with $H_3(Z_{{i_1}{i_2}}^l)=0.5$ bits. We consider three broadcast messages as follows: (i) $Z_{12}^1$, (ii) $Z_{23}^2$, and (iii) $Z_{13}^1 \oplus_3 Z_{13}^2$. From (i) and (iii), $f_1$ is decoded; using (i) and (ii) $f_2$ is decoded; and finally, $f_3$ is decoded given (ii) and (iii). Thus, $R_{M}$ for broadcasting (i)-(iii), exploiting the independence bound on entropy, satisfies\footnote{Given a length $n$ sequence $X_j^n$, we denote two disjoint splits of $X_j^n$ by $X_j^{{n}/{2}} = \{X_{j,1}, \dots, X_{j,{n}/{2}}\}$ and $X_{j}^{{n}/{2}+1:n}=\{X_{j,{n}/{2}+1}, \dots, X_{j,n}\}$ two sub-sequences of length $\frac{n}{2}$. Combining the idea of splitting with the above-mentioned compatible function construction enables the asymptotically lossless reconstruction of $\{f_i^n: i\in[3]\}$ for sufficiently large $n$.} 
 \begin{align}
 \label{ex-lineaer-rate-split}
 R_M&\leq H_3(Z_{12}^{1},Z_{23}^{2},Z_{13}^{1}\oplus_{3}Z_{13}^{2})\nonumber\\
 &\leq H_3(Z_{12}^{1})+H_{3}(Z_{23}^{2})+H_{3}(Z_{13}^{1}\oplus_{3}Z_{13}^{2})=1.5\:\text{ bits} .\nonumber
 \end{align}

{\bf Vector-based approach.}   
We next propose a vector-based achievability scheme, drawing on the notion of characteristic graphs, that involves designing two messages, $M_{1}$ and $\textbf{M}_{2}$, derived from $W\in\Gamma(G_{\cup_{i=1}^{3}f_{i}})$. When $X_{2}=0$, with a probability $1/3$, the MISs are determined by the function $X_{1}\oplus_{3}X_{3}$ and encode these MISs to the message $M_{1}$ with a rate $H_3(M_{1})=H_3(X_{1}\oplus_3 X_{3})=1$. Similarly, when $X_{2}\in [2]$, with a probability $2/3$, the MISs are determined by $X_1\oplus_{3}X_{2}\oplus_{3} X_{3}$ and $X_{2}$. These MISs are encoded to message $\textbf{M}_{2}$ with a rate of $H_3(\textbf{M}_{2})=H_{3}(X_1\oplus_3 X_{2}\oplus_{3} X_{3}, X_{2})=1.63$ bits. Thus, the rate required to broadcast these messages is
\begin{equation}
    R_M\leq\frac{1}{3}H_{3}(M_{1})+\frac{2}{3}H_{3}(\textbf{M}_{2})=1.42\: \text{bits} \ ,
\end{equation}
compared to the achievable rate of $1.5$ bits in \cite[Example~2]{yao2024capacity}.
\end{ex}

Example~\ref{ex-linear-comp-jafar} presents two approaches for linear function demands: a compatibility function-based technique (similar to~\cite{yao2024capacity}) and a vector-based technique. 
The performance of the former approach is similar to the linear coding scheme of~\cite[Example~2]{yao2024capacity} with a rate of $1.5$ bits. The latter 
showed that by carefully designing the broadcast message based on the nature of the demands and the datasets, we can achieve $R_{M}\leq 1.42$ bits, indicating an improvement over the compatibility function-based approach that completely relies on the linear encoding of the demands. 
This shows that a vector-based approach, refining the characteristic graph-based broadcast message design, can surpass the state of the art. However, achieving the minimum rate heavily depends on the structure of the demanded function and lacks a generalizable form. 

The next section presents our achievable and converse rates for the $K$-user computation broadcast model.

\section{\texorpdfstring{$K$}{}-User Computation Broadcast: Achievability and Converse Bounds}
\label{sec:main-resu}
This section presents an upper bound $R_{ach}$ and a lower bound $R_{con}$ on the communication rate for the $K$-user computation broadcast model, as shown in Figure~\ref{fig:CBNs}.

\begin{theo}
\label{theo-inner-and-outer}
    Given a $K$-user computation broadcast model, with $N$ datasets $X_{[N]}$, and a  broadcast graph $G_{\cup_{i=1}^{K}f_i}$ the communication rate for the broadcast message $M$ satisfies:
    \begin{align}
    \label{eq-theo-inner-and-outer}
        R_{ach}= &\min_{X_{[N]}\in W\in \Gamma(G_{\cup_{i'=1}^{K}f_{i'}})}\max_{i\in [K]}I(X_{[N]};W\mid \mathcal{S}_{i}) \ ,\\
        R_{con}=&\min_{p(u\mid x_{[N]})}\max_{i\in [K]}I(X_{[N]},U\mid \mathcal{S}_{i}) \ ,
    \end{align}
    where $\Gamma(G_{\cup_{i=1}^{K}f_i})$ is the set of MISs of $G_{\cup_{i=1}^{K}f_i}$.
\end{theo}

\begin{proof}
\label{proof-Theorem-upper-and-lower} 
To derive $R_{ach}$, we here use the compress-bin strategy~\cite[Theorem 11.3]{el2011network}. 
Consider length $n$ i.i.d realizations of all sources, i.e., $X_{[N]}^{n}=\{X_{1,t},\dots,X_{N,t}\}_{t\in [n]}$, drawn from $P_{X_{[N]}}$. 
Define $\hat{f}_{i}^{n}=F_i(W^n, \mathcal{S}_{i}^{n})$, denoting an estimate of $f_{i}^{n}$ at user $i\in [K]$, 
with $W^n$ denoting an MIS of $G^n_{\cup_{i=1}^{K} f_i}$ built using the length $n$ realizations $X_{[N]}^{n}$ and $\{\mathcal{S}_{i}^{n}\}_{i\in[K]}$.

\paragraph{Codebook generation}  
The master node generates $q^{n\Rcomp}$ i.i.d sequences $W^{n}(l)$, where $l \in [0, q^{n\Rcomp} - 1]$. For $\Rcomp>R_{ach}$, the index set $[0, q^{n\Rcomp}-1]$ is uniformly partitioned into $q^{nR_{ach}}$ bins, $\mathcal{B}(m)=[(m-1)q^{n(\Rcomp-R_{ach})}: mq^{n(\Rcomp-R_{ach})}-1]$, for $m\in [1, q^{nR_{ach}}]$. The codebook, denoted as $\mathcal{C}=\{m\in [1,q^{nR_{ach}}]\}$, is revealed to all users.

\paragraph{Encoding}
Given $X_{[N]}^{n}$, the master node finds an index $l^*\in [1:q^{n\Rcomp}-1]$ such that $(W^{n}(l^*),X^{n}_{[N]})\in \mathcal{T}_{\epsilon'}^{(n)}$, where $\mathcal{T}_{\epsilon'}^{(n)}$ represents $\epsilon'$-typical sequences $(W^n, X^{n}_{[N]})$ for some $\epsilon'>0$. 
If there are multiple such indices, the master randomly selects one of them. If no such index exists, the master selects $l^*=0$. The master node broadcasts $m^*$ such that $l^*\in \mathcal{B}(m^*)$. 
Thus, $(l^*, m^*)$ represents the selected indices at the master.

\paragraph{Decoding}
Let $\epsilon>\epsilon'$. User $i\in [K]$ receives $m^*$, and finds the unique index $\hat{l}_i \in \mathcal{B}(m^*)$, such that $(W^{n}(\hat{l}_i),\mathcal{S}_{i}^{n})\in \mathcal{T}_{\epsilon}^{(n)}$. User $i\in [K]$ then computes $F_i(W^n(\hat{l}_i), \mathcal{S}_{i}^{n})$. 
It declares an error if there is no such $\hat{l}_i$.

\paragraph{Analysis of expected error}
Let $W^{n}(\hat{l}_i)$ be the message decoded by user $i\in[K]$. An error event at user $i\in[K]$ is denoted as $\err_{i}=\{(W^{n}({\hat{l}_i}), X_{[N]}^{n})\notin \mathcal{T}_{\epsilon}^{(n)}\}=\{\hat{l}_i\ne l^*\}$, which can be categorized into three events:
\begin{align}
    \err_{i1}=&\{(W^{n}(l), X_{[N]}^{n})\notin \mathcal{T}_{\epsilon'}^{(n)}, \text{ for all } l \in [0:q^{n\Rcomp}-1] \} \ ,\nonumber\\
    \err_{i2}=&\{(W^n(l^*), X_{[N]}^{n}
    )\notin \mathcal{T}_{\epsilon}^{(n)}\} \ , \nonumber\\
    \err_{i3}=&\{(W^{n}(l), \mathcal{S}_{i}^{n})\in \mathcal{T}_{\epsilon}^{(n)}, \text{ for some } 
    l\in \mathcal{B}(m^*), \:l\ne l^* \} \ .\nonumber
\end{align}
As $n\to\infty$, by covering lemma (cf.~\cite[Lemma~3.3]{el2011network}), $P(\err_{i1})\to 0$ if $\Rcomp> I(X_{[N]}; W)+\delta(\epsilon')$, where $\delta(\epsilon')>0$ tends to zero as $\epsilon'\to 0$, by the conditional typically lemma~\cite[Section~2.5]{el2011network} $P(\err_{i2} \cap \err_{i1}^c)\to 0$, and  by packing lemma \cite[Lemma~3.1]{el2011network}, $P(\err_{i3})\to 0$ if $\Rcomp-R_{ach}< I(W;\mathcal{S}_{i})-\delta(\epsilon)$. Using these lemmas yields  $R_{ach}>I(X_{[N]};W\mid \mathcal{S}_{i})$ for any $i\in[K]$. Hence, it must hold that
\begin{equation}
\label{eq:rate_condition}
    R_{ach}\geq \max_{i \in [K]} I(X_{[N]};W\mid \mathcal{S}_{i})\ ,
\end{equation}
where similar techniques exist for lossy source coding with side information, see e.g., the achievability scheme of~\cite{heegard1985rate}.

Next, we derive the lower bound $R_{con}$. Given $M$, 
user $i\in [K]$ computes $\hat{f}_{i}^{n}$ upon receiving $M$ and $\mathcal{S}_{i}^{n}$. 
The $t$-th realization $f_{i,t}$, for $t\in [n]$, is defined as $f_{i,t}=g_{i,t}(U_{t}, \mathcal{S}_{i,t})$ for some function $g_{i,t}$ where $U_{t}$ is an  auxiliary variable defined as $U_{t}=(M, \mathcal{S}_{i}^{t-1})$. Letting $t=n$, we obtain 
\begin{align*}
    nR_{ach}\geq& H_q(M) \overset{(a)}{\geq}H_q(M\mid \mathcal{S}_{i}^{n}) \overset{(b)}{=}I(X_{[N]}^{n}; M\mid \mathcal{S}_{i}^{n})\ , \\
    &\overset{(c)}{=}\sum_{t=1}^{n}I(X_{[N],t};M\mid \mathcal{S}_{i}^{n}, X_{[N]}^{t-1}) \ ,\\
    &\overset{(d)}{=}\sum_{t=1}^{n}I(X_{[N],t};M,X_{[N]}^{t-1}, \mathcal{S}_{i}^{t-1}\mid \mathcal{S}_{i,t} ) \\ 
    &\overset{(e)}{=}\sum_{t=1}^{n}I(X_{[N],t};U_{t}\mid \mathcal{S}_{i,t})\ ,
\end{align*}
where $(a)$ follows from conditioning, $(b)$  from noting that $M$ is function of $X_{[N]}^{n}$, i.e., $H_q(M\mid \mathcal{S}_{i}^{n},X_{[N]}^{n})=0$, $(c)$ from the chain rule of mutual information, $(d)$ is due to the independence of $X_{[N],t}$ from $(X_{[N]}^{t-1}, \mathcal{S}_{i}^{t-1},\mathcal{S}_{i, t+1}^{n})$, and $(e)$ from setting the auxiliary variable $U_{t}=(M,X_{[N]}^{t-1}, \mathcal{S}_{i}^{t-1})$.
\end{proof}

We next introduce a tight converse bound, different from $R_{con}$ in Theorem~\ref{theo-inner-and-outer},
derived using the joint PMF of each of the demanded functions and the available side information.

\begin{prop}
\label{prop-lower-bound-mutual-info}    
The lower bound on the communication rate for satisfying all users' demands, given side information $\mathcal{S}_i$ is
\begin{align}
\label{eq-prop-lower-bound-mutual-info} 
H_q(f'_1, f_2', \cdots f_K') \ ,
\end{align}
where $f_i'= f_i|\mathcal{S}_i$ is a conditional random variable capturing the demand $f_i$ of user $i\in [K]$ given $\mathcal{S}_i$, where $f_i' \in \mathbb{F}_q$. 
\end{prop}

\begin{proof}
Assume that the master generates a broadcast message $M$ that satisfies (\ref{Recovery_condition}) for all users. Since each user $i\in [K]$ can calculate $f_i'$ from the received message, we have
\begin{align}
\label{eq-proof-prop-lower-bound-mutual-info}
    H_q(f_1',f_2', \cdots, f_K'\mid M)=0 \ .
\end{align}
Next, consider the mutual information between $M$ and $f_i'$,
\begin{align}
\label{eq-2-proof-prop-lower-bound-mutual-info-1}
    I(f_1', \cdots, f_K'; M)&=H_q(f_1', \cdots, f_K')-H_q(f_1', \cdots, f_K'| M), \nonumber \\ 
    \overset{(a)}{=}&H_q(f_1', \cdots, f_K')\ ,
\end{align}
where $(a)$ follows from (\ref{eq-proof-prop-lower-bound-mutual-info}). By expanding (\ref{eq-2-proof-prop-lower-bound-mutual-info-1}) we have
\begin{align}
\label{eq-2-proof-prop-lower-bound-mutual-info-2}
    I(f_1', \cdots, f_K'; M)=H_q(M)-H_q(M\mid f_1', \cdots, f_K')\ .
\end{align}
From (\ref{eq-2-proof-prop-lower-bound-mutual-info-1}) and (\ref{eq-2-proof-prop-lower-bound-mutual-info-2}), $H_q(f_1', \cdots, f_K')\leq H_q(M)\leq R$.
\end{proof}

For comparison, we generalize the genie-aided lower bound in~\cite{jafar2006capacity} to our model, as detailed next.

\begin{prop}
\label{prop-genie}  
(A genie-aided lower bound~\cite{jafar2006capacity} tailored to non-linear computation broadcast model.) 
Given a $K$-user computation broadcast network, where each user requests $f_i$ for $i \in [K]$, a lower bound on the broadcast rate is given as follows:
\begin{align}
\label{eq-geini-aided-rate}
R_{Genie}\geq&\sum_{i=1}^{K}H_q(f_i \mid \mathcal{S}_{[i-1]},\mathcal{S}_{i},\{f_{i'}\}_{i'\in[i-1]}) \ ,
\end{align}
where given the 
set of users in $[i-1]$ with $i\in\{2,\dots, K\}$, a genie provides the side information $\{\mathcal{S}_{i'}\}_{i'\in[i-1]}$ and function demands $\{f_{i'}\}_{i'\in[i-1]}$ to user $i$.  
\end{prop}

While the approach in Proposition~\ref{prop-lower-bound-mutual-info} satisfies the demands of $K$ users simultaneously by generating a broadcast message without allowing any interaction between the $K$ users, 
Proposition~\ref{prop-genie} incorporates such interactions via extending the genie-aided approach devised for two users in~\cite{jafar2006capacity}, to the $K$ user setting. 
To that end, we employ a successive refinement-based method for recovering the demands, which requires users to share side information as detailed in Proposition~\ref{prop-genie} that yields the rate lower bound in (\ref{eq-geini-aided-rate}).

To demonstrate the applicability of the lower bound in (\ref{eq-geini-aided-rate}), consider the $3$-user Boolean computation broadcast problem in Example~\ref{ex-nonlinear-comp}. The lower bound, using (\ref{eq-geini-aided-rate}), is given as:
 \begin{align}
 \label{eq-low-bound-gein-ex-3-user-nonlin}
 R_{Genie}\geq &H_2(f_3|X_3)+H_2(f_1|X_1, X_3, f_3) \nonumber\\  
+& H_2(f_2|X_2,X_1,X_3, f_1,f_3) \nonumber\\
=&0.905+0.25+0=1.155\: \text{ bits}\ .
 \end{align}

The application of the lower bound (\ref{eq-prop-lower-bound-mutual-info}) in Proposition~\ref{prop-lower-bound-mutual-info} to the $3$-user Boolean computation model of Example~\ref{ex-nonlinear-comp} yields $1.45$ bits, 
while the achievable rate, from (\ref{R-3-user}), is $R_{ach}=1.5$ bits, indicating a tight gap between the bounds. 
This comparison shows that (\ref{eq-prop-lower-bound-mutual-info}) in Proposition~\ref{prop-lower-bound-mutual-info} yields a tighter lower bound than $R_{Genie}\geq 1.155$ bits in (\ref{eq-geini-aided-rate}) in Proposition~\ref{prop-genie}, which is a generalization of the genie-aided lower bound in~\cite{jafar2006capacity}, for serving this type of non-linear demands.

{\bf Discussion.} 
In this paper, we examined the $K$-user computation broadcast model with general data distributions and general function demands over finite fields.
Using the characteristic graph framework, we proposed a novel broadcast coding scheme that reduces the communication rate while accommodating both linear and non-linear demands compared to the state of the art. Moreover, we presented achievable and converse rates for the general $K$-user computation broadcast problem, while more research is needed to quantify the gap between the bounds. 
Future directions include exploring more intelligent storing strategies to reduce communication complexity, where available side information may consist of functions of master's datasets (instead of pure datasets), and where users might have heterogeneous storage capacities.


\Urlmuskip=0mu plus 1mu\relax
\bibliographystyle{IEEEtran}
\bibliography{ref}

\end{document}